\documentclass[preprint]{vgtc}               

\ifpdf
  \pdfoutput=1\relax                   
  \usepackage{graphicx}                
  \DeclareGraphicsExtensions{.pdf,.png,.jpg,.jpeg} 
\else
  \usepackage{graphicx}                
  \DeclareGraphicsExtensions{.eps}     
\fi%

\graphicspath{{figures/}{pictures/}{images/}{./}} 

\usepackage{microtype}                 
\PassOptionsToPackage{warn}{textcomp}  
\usepackage{textcomp}                  
\usepackage{mathptmx}                  
\usepackage{times}                     
\usepackage{cite}                      
\usepackage{tabu}                      
\usepackage{booktabs}                  

\onlineid{0}

\vgtccategory{Research}

\vgtcinsertpkg

\title{Video Analysis of Behavioral Patterns During Prolonged Work in VR}

\author{
 Verena Biener$^{1}$\thanks{e-mail: verena.biener@hs-coburg.de}
 \and Forouzan Farzinnejad$^{1}$
\and Rinaldo Schuster$^{1}$
\and Seyedmasih Tabaei$^{1}$
\and Leon Lindlein$^{1}$
\and Jinghui Hu$^{2}$
\and Negar Nouri$^{1}$
\and John J. Dudley$^{2}$
\and Per Ola Kristensson$^{2}$
\and Jörg Müller$^{3}$
\and Jens Grubert$^{1}$\thanks{e-mail: jens.grubert@hs-coburg.de}
}
\affiliation{\scriptsize $^{1}$Coburg University of Applied Sciences and Arts, Germany\\  $^{2}$University of Cambridge, United Kingdom \\  $^{3}$University of Bayreuth, Germany}

\teaser{
  \centering
  \includegraphics[width=\linewidth]{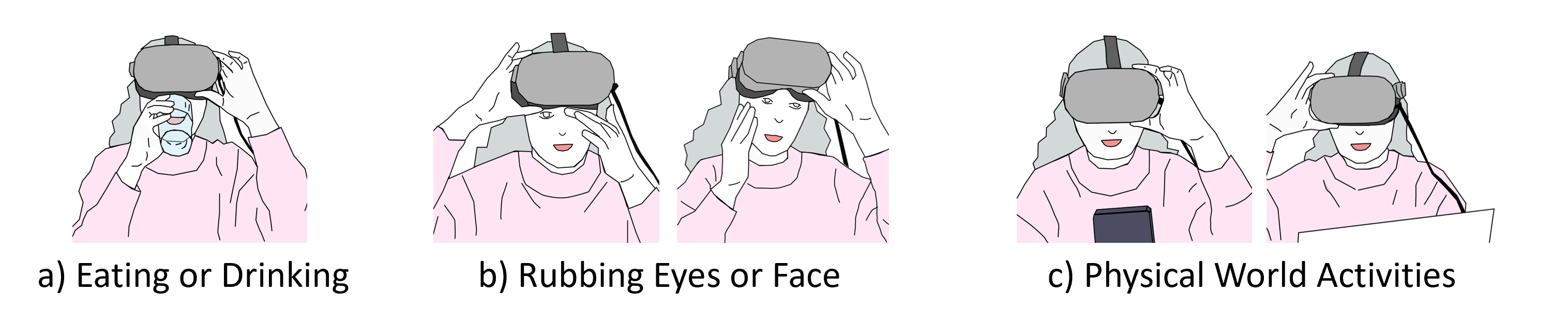}
  \caption{Actions that were observed during the video annotation process. (a) Eating or drinking; (b) Rubbing eyes or face; (c) Interacting with the physical world objects, like phones or papers. These actions were observed less frequently in \textsc{vr}.}
  \label{fig:teaser}
}

\abstract{VR has recently been promoted as a tool for knowledge workers and studies have shown that it has the potential to improve knowledge work. However, studies on its prolonged use have been scarce. A prior study compared working in VR for one week to working in a physical environment, focusing on performance measures and subjective feedback. However, a nuanced understanding and comparison of participants' behavior in VR and the physical environment is still missing. To this end, we analyzed video material made available from this previously conducted experiment, carried out over a working week, and present our findings on comparing the behavior of participants while working in VR and in a physical environment.
} 


\begin{document}


\firstsection{Introduction}

\maketitle

The opportunities for Virtual Reality (VR) as a medium for knowledge work, as opposed to entertainment, are beginning to be explored~\cite{biener2020breaking,ruvimova2020transport,biener2022povrpoint}.
VR has the potential to enhance the working experience through multiple mechanisms, such as exploiting large virtual displays~\cite{biener2020breaking, mcgill2020expanding}, personalizing the work-environment~\cite{ruvimova2020transport}, or delivering enhanced interactivity~\cite{biener2022povrpoint}.
However, for anyone who has worn a current-generation VR headset for an extended period of time, the notion of completing even a single work day wearing a head-mounted display (HMD) could be a daunting prospect. 

The demands of knowledge work, in contrast to VR in gaming or leisure, suggest that workers may either willingly or unwillingly spend extended periods wearing an HMD without the welcome distraction of entertainment. 
Recognition of this fact motivates research seeking to understand how knowledge workers respond to extended use of VR \cite{guo2020exploring, shen2019mental}.
So far the longest study was conducted by Biener et al.~\cite{biener2022quantifying} in which participants were completing a full work week in VR while doing their normal work tasks.
This was then compared to a week in which participants worked in a comparable physical setup.
They found that the VR condition delivered significantly worse ratings across measures of task load, frustration, negative affect, anxiety, eye strain, system usability, flow, productivity, wellbeing and simulator sickness.
Nevertheless, some of the reported measures slightly improved over the five days.
While conducting this study, Biener et al.~\cite{biener2022quantifying} amassed a rich dataset of over 1,400 hours of video, capturing participant behavior in both conditions throughout both weeks, but this was not analysed in the original paper.
We used this dataset 
and analyzed videos of one day per condition for all 16 participants to report on their behavior as they respond to the experience of working in VR as compared to working in the physical environment.
To our knowledge, this is the first study of its kind and delivers substantial insight into the, as yet, undocumented behavior of users in a novel VR work setting.


\section{Methodology}
The goal of this work was to closely analyze the behavior of 16 participants (6 female, 10 male, mean age $=29.31$ years, $sd=5.52$) during an experiment involving prolonged use of VR in a work setting. The participants were observed wearing HMDs during their work for five continuous days (\textsc{vr} condition) and then for another five days without an HMD (\textsc{physical} condition), as described in Biener et al.~\cite{biener2022quantifying}. Each day, they worked for eight hours with a 45 minute lunch break after four hours.
The videos we use here were recorded during that study~\cite{biener2022quantifying} using a webcam.
We analyzed the participants' behavior, by involving six people in watching and annotating interesting behavior in the videos (Annotator A, B, C, D, E, and F).
Through an iterative process, annotator A and B created a codebook for labelling behavior in \textsc{vr}.
The codebook for the \textsc{physical} videos was derived as closely as possible from the \textsc{vr} codebook, allowing a comparison of the behavior from \textsc{vr} with the standard behavior in \textsc{physical}. 
The other annotators were then trained by annotator A and B and samples of annotations were compared between annotators.
To obtain more consistency, all videos of one participant were only worked on by one annotator.

Due to the high demand on time for labeling one video (about one hour for processing one hour of video material),
we report only on the comparison of the fifth day in \textsc{vr} and the first day of \textsc{physical}. This was chosen, because participants would be most familiar with the HMD on their last day and because participants were already familiar with working with a standard desktop setup, so we do not expect a change of behavior over time for the \textsc{physical} videos.

\section{Results}
We analyzed the occurrence of different categories of events by comparing \textsc{day 5} of \textsc{vr} to \textsc{day 1} of \textsc{physical} using a repeated measures ANOVA with the independent variable \textsc{interface} (\textsc{vr}, \textsc{physical}) and \textsc{time} (\textsc{morning}, \textsc{afternoon}), similar as in prior work~\cite{biener2022quantifying}.
To get the average per hour, we divided the total number and duration of events during each time-period (\textsc{morning}, \textsc{afternoon}) by the duration of this time-period, which was usually around four hours and 22 minutes (four hours of work plus half of the 45 minute break).
Whenever the sphericity assumption was violated, we used Greenhouse-Geisser correction to ensure the robustness of the ANOVA~\cite{blanca2023non}.
We applied Bonferroni-correction to all post-hoc tests involving multiple comparisons.



\paragraph{\textbf{Screen Time:}}
We calculated screen time in \textsc{vr} by considering all times in which participants were wearing the HMD, not including times in which they were taking the HMD halfway off, using the controller, having problems with the keyboard tracking or were reading, writing or using their phone.
In \textsc{physical} we labelled all times in which the participants were facing the screen, also not including times in which they were reading, writing or using the phone. This approach could not be reliably used for \textsc{vr}, because it is not clear from the videos if participants were looking at the virtual monitor.
Comparing \textsc{vr} to \textsc{physical} showed no main effect of \textsc{interface}, suggesting that the time participants spent working was not significantly different between conditions.


\paragraph{\textbf{Standing up:}}
We labelled all parts where participants were standing up or sitting down.
Comparing \textsc{vr} to \textsc{physical} showed no significant difference in the number of standing events, however, we found a significant influence of \textsc{time} ($F(1,15)=66.97$, $p<0.001$, $\eta^2_p=0.82$) on the duration of standing per hour and no interaction effect between \textsc{interface} and \textsc{time}. 
Post-hoc tests indicated that both in \textsc{vr} and \textsc{physical} participants stood significantly less ($p<0.001$) during the \textsc{afternoon} (\textsc{vr}: $m=734.08~sec$, $sd=438.32$; \textsc{physical}: $m=750.38~sec$, $sd=594.27$) than during the \textsc{morning} (\textsc{vr}: $m=3605.99~sec$, $sd=1459.87$; \textsc{physical}: $m=3244.79~sec$, $sd=1561.97$).

\paragraph{\textbf{Eating or Drinking:}}
This category describes all parts of the video in which participants put drinks or food into their mouth, such as displayed in Fig. \ref{fig:teaser} (a), not including longer periods of chewing.
Comparing \textsc{physical} to \textsc{vr}, we found that the \textsc{interface} had a significant influence on the number of such events per hour ($F(1,15)=7.97$, $p=0.013$, $\eta^2_p=0.45$), such that in \textsc{physical} ($m=3.19$, $sd=2.67$) participants were drinking or eating more often than in \textsc{vr} ($m=1.79$, $sd=1.31$). 
However, we could not find a significant influence on the time spent eating or drinking per hour.
This suggests that participants took longer to eat or drink in VR which could be caused by them being more cautious.

\paragraph{\textbf{Rubbing Eyes or Face:}}
This category was used whenever participants were rubbing their eyes or face, as illustrated in Fig. \ref{fig:teaser} (b).
When comparing \textsc{vr} to \textsc{physical}, we found a significant main effect of \textsc{interface} ($F(1,15)=5.6$, $p=0.032$, $\eta^2_p=0.27$) on the number of such actions, indicating that participants rubbed their eyes and faces more in \textsc{physical} ($m=2.87$, $sd=2.25$) than \textsc{vr} ($m=1.68$, $sd=1.44$).
We also found significant main effects of \textsc{time} ($F(1,15)=7.64$, $p=0.014$, $\eta^2_p=0.34$)
indicating more such actions in the \textsc{morning} ($m=2.75$, $sd=2.36$) compared to the \textsc{afternoon} ($m=1.8$, $sd=1.36$). However, post-hoc tests did not confirm this for \textsc{vr} and \textsc{physical} individually. 
We also did not find a significant effect of \textsc{interface} on the total time spent on such actions.

\paragraph{\textbf{Physical World Activities:}}
We were also interested in how often and for how long participants concerned themselves with things outside of the virtual world, as depicted in Fig. \ref{fig:teaser} (c). Therefore, this category includes events where the annotators believe the participants were reading or writing something outside of VR, using a smartphone, or otherwise peeking under the HMD.
Comparing \textsc{vr} to \textsc{physical} showed a significant main effect of \textsc{interface} ($F(1,15)=7.669$, $p=0.014$, $\eta^2_p=0.34$) such that there are significantly more such actions in \textsc{physical} ($m=6.88$, $sd=8.40$) than in \textsc{vr} ($m=3.28$, $sd=4.97$), but no significant differences have been found regarding the total time spent on such actions.

\section{Discussion and Conclusion}
Analyzing the videos showed that some actions were more common in \textsc{physical} as compared to \textsc{vr} such as consuming food, rubbing faces and eyes or interacting with the physical world. However, no significant difference could be found for the total duration of such events per hour. An explanation for this could be that it takes more effort to perform such actions in \textsc{vr} and therefore participants do them less often, but for prolonged times, or more cautiously. Also in \textsc{physical} some of these actions could happen involuntarily.
For other actions (screen time, standing up), no significant difference could be found between \textsc{physical} and \textsc{vr}, on the contrary, similar behavior was detected such as less time spent standing in the afternoon.

These results already provide interesting insights into the behavior of users while working in VR for a much longer timescale than the duration of many VR user studies so far. However, the videos recorded in the prior study \cite{biener2022quantifying} can provide much more insights into user behavior than presented in the scope of this paper, such as the changes in behavior over the period of a full work week and detailed observations that are only accessible through a video analysis. Therefore, we are planning to analyze and report on such more in-depth findings in future work.


\bibliographystyle{abbrv-doi}

\bibliography{template}
\end{document}